\documentclass[aps,prb,twocolumn,showpacs,tightenlines]{revtex4}
\usepackage{graphics}

\begin{document}

\title{Cotunneling thermopower of single electron transistors}

\author{M. Turek} 

\altaffiliation[Present address: ]{University of Regensburg, Institute for
  theoretical physics, D-93040 Regensburg, Germany}

\author{K.A. Matveev} 
\affiliation{Department of Physics, Duke University, Durham, NC 27708-0305}

\date{\today} 

\begin{abstract}
  We study the thermopower of a quantum dot weakly coupled to two
  reservoirs by tunnel junctions.  At low temperatures the transport
  through the dot is suppressed by charging effects (Coulomb blockade).
  As a result the thermopower shows an oscillatory dependence on the gate
  voltage.  We study this dependence in the limit of low temperatures
  where the transport through the dot is dominated by the processes of
  inelastic cotunneling.  We also obtain a crossover formula for
  intermediate temperatures which connects our cotunneling results to the
  known sawtooth behavior in the sequential tunneling regime.  As the
  temperature is lowered, the amplitude of thermopower oscillations
  increases, and their shape changes qualitatively.
\end{abstract}

\pacs{73.23.Hk, 73.50.Lw, 72.15.Jf}

\maketitle

\section{Introduction}

\begin{figure}[b]
 \resizebox{.38\textwidth}{!}{\includegraphics{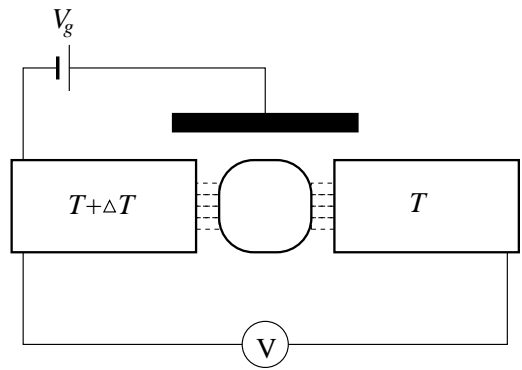}}
\caption{\label{system} 
  Thermopower measurement in a single electron transistor.  The quantum
  dot is coupled to the two leads by tunnel barriers.  The electrostatics
  of the system is controlled by the voltage $V_g$ applied to a gate
  coupled capacitively to the quantum dot.  The left and right leads are
  held at temperatures $T+\Delta T$ and $T$, respectively.  As a result a
  small voltage $V$ is generated between the leads.  The resulting
  thermopower $S=-V/\Delta T$ is measured as a function of $V_g$.}
\end{figure}

In the last few years a number of experiments have been performed in order
to investigate the transport of electrons through small conductors, such
as metallic particles and semiconductor quantum dots.
\cite{Grabert,Kouwenhoven}  One of the most commonly studied types of devices
is the {\em single-electron transistor\/} which consists of a quantum dot (or
a small metal particle) connected to two leads by tunnel junctions.  The
particle is usually also capacitively coupled to an additional gate
electrode, Fig.~\ref{system}.  The transport of electrons through the
quantum dot is strongly affected by charging effects.  Indeed, when an
electron tunnels into the dot from a lead, the electrostatic energy of the
system increases by $\sim E_C \equiv e^2/2C$, 
where $e$ is the elementary charge, and
$C$ is the capacitance of the dot.  In a typical experiment the
temperature is low, $T\ll E_C$, and the tunneling is strongly
suppressed as only a very small fraction of electrons have energy of the
order of $E_C$ necessary for the tunneling to occur.  This phenomenon
is commonly referred to as the Coulomb blockade.  In a single electron
transistor, the charging energy can be controlled by the gate voltage
$V_g$.  For instance, by applying a positive voltage to the gate, one can
lower the increase in electrostatic energy caused by adding an extra
electron to the dot.  As a result at certain values of $V_g$ the
electrostatic gap vanishes, and the transport of electrons through the
system is strongly enhanced.  This leads to a sequence of periodic peaks
in the conductance of the single electron transistor as a function of
$V_g$, which is often observed in the experiments. \cite{Kouwenhoven}  The
positions and the shape of the peaks are in good agreement with the
so-called sequential tunneling theory \cite{glazman:G,beenakker:G} of
conductance which accounts for real processes of electron tunneling
between the dot and the leads.

In a number of more recent experiments the thermopower $S$ of a single
electron transistor has been studied.
\cite{staring:S,dzurak:S,dzurak:exp,Molenkamp98} Similarly to the case of
conductance, the thermopower shows periodic oscillations as a function of
the gate voltage.  The theory of the thermopower in a single electron
transistor in the framework of the sequential tunneling approach was
developed by Beenakker and Staring.  \cite{beenakker:S} It was confirmed
by the experimental data of Ref.~\onlinecite{staring:S}, where the
sawtooth shape of the thermopower oscillations predicted in
Ref.~\onlinecite{beenakker:S} was observed.  On the other hand, in a
recent experiment \cite{dzurak:exp} the observed thermopower oscillations
had a shape different from the sawtooth, and also the amplitude of the
oscillations was far smaller than the one predicted in
Ref.~\onlinecite{beenakker:S}.

The most significant difference between the experiments of
Ref.~\onlinecite{staring:S} and Ref.~\onlinecite{dzurak:exp} was that the
temperature in the latter work was very small, $T\sim 0.006\ e^2/C$, in
comparison with the estimated temperature $T\sim 0.065\ e^2/C$ in
Ref.~\onlinecite{staring:S}.  The authors of Ref.~\onlinecite{dzurak:exp}
attributed the deviation of their data from the theory \cite{beenakker:S}
to the fact that the theory neglected the effects of virtual tunneling
({\it cotunneling}) of electrons through the dot.  Indeed, it is known
that in the case of very low temperatures cotunneling processes
\cite{averin:cot} give dominant contribution to the conductance of single
electron transistors.  One can therefore expect that this mechanism will
result in a different behavior of the thermopower at low temperatures.

In this paper we develop the theory of the thermopower of single electron
transistors in the regime of inelastic cotunneling.  This mechanism
dominates the low temperature electron transport in the case of relatively
large dots, where the effects of finite quantum level spacing can be
neglected.  We find the thermopower oscillations of the shape
qualitatively similar to that observed in the experiment,
\cite{dzurak:exp} and the amplitude of the oscillations is significantly
lower than the result of the sequential tunneling theory.
\cite{beenakker:S} We also study the crossover from the cotunneling
behavior of the thermopower to the sawtooth regime of sequential tunneling
which occurs as one raises the temperature $T$ of the system above a
certain crossover temperature $T_c$.  The latter is found to be $T_c
\simeq E_C / \ln[e^2/\hbar(G_l+G_r)]$, where $G_l$ and $G_r$ are the
conductances of the tunneling junctions connecting the quantum dot to the
left and right leads.  Throughout the paper $G_l$ and $G_r$ are assumed to
be small, $G_l,G_r\ll e^2/\hbar$.  The opposite regime of strong
tunneling, more relevant for the conditions of the
experiment,\cite{Molenkamp98} was recently addressed in
Ref.~\onlinecite{Andreev}.

In the next section we introduce the theoretical model of a single
electron transistor used in this paper and discuss the relevant mechanisms
of electronic transport in the device. In Sec.~\ref{thermopower} we review
the known results for the thermopower $S$ in the regime of sequential
tunneling and obtain $S$ in the regime of inelastic cotunneling.  In
Sec.~\ref{general} these results are unified in a single formula that
correctly describes the crossover between the two regimes.

\section{Mechanisms of transport\label{mechanisms}}

To describe the single electron transistor Fig.~\ref{system}, we introduce 
the Hamiltonian in the form $\hat{H}=\hat{H_0}+\hat{V}$, where
\begin{eqnarray}
\label{sys}
 \hat{H_0} &=& \sum_k \xi_k a_k^\dagger a_k + \sum_p \xi_p a_p^\dagger a_p 
         +\sum_q \xi_q a_q^\dagger a_q  \nonumber\\
           &&+ \frac{\hat{Q}^2}{2 C} + \phi \hat{Q},\\
 \hat{V} &=& \sum_{k,p_1} \left(t_{k,p_1} a_k^\dagger a_{p_1} + 
          t_{k,p_1}^* a_{p_1}^\dagger a_k\right) \nonumber\\
         &&+ \sum_{p_2,q} \left(t_{p_2,q} a_{p_2}^\dagger a_q + 
          t_{p_2,q}^* a_q^\dagger a_{p_2}\right), \\
 \hat{Q} &=& -e \sum_p \left[a_p^\dagger a_p - \theta(-\xi_p)\right].
\end{eqnarray}
Here $a_k$, $a_p$, and $a_q$ are the annihilation operators for the
electrons in the left lead, quantum dot, and the right lead, respectively;
the electron energies $\xi_k$, $\xi_p$, and $\xi_q$ are measured from the
Fermi level; the tunneling in and out of the dot is described by matrix
elements $t_{k,p_1}$ and $t_{p_2,q}$.  Operator $\hat Q$ represents the
charge of the quantum dot, $e$ is the elementary charge, and the potential
$\phi$ is proportional to the gate voltage $V_g$.

Our goal is to find the thermopower $S$ of the system as a function of the
potential $\phi$ for a given temperature.  The thermopower is defined in
terms of the voltage $V$ generated across the single electron transistor
when the temperatures of the left and right leads $T_l$ and $T_r$ differ
by $\Delta T\ll T_l,T_r$ and no current $I$ through the system is allowed:
\begin{equation}
  \label{eq:S_definition}
   S\equiv - \left. \lim_{\Delta T \to 0} \frac{V}{\Delta T}
   \right|_{I=0}.
\end{equation}
In the linear response regime the current can be presented as 
\begin{equation}
  \label{eq:linear_response}
  I= G_T \Delta T + G V,
\end{equation}
where $G$ is the conductance of the system, and the kinetic coefficient
$G_T$ describes the current response to the temperature difference. 
The condition $I=0$ then results in the following expression for the
thermopower: 
\begin{equation}
  \label{S}
  S = \frac{G_T}{G}.
\end{equation}
Thus, one can find the thermopower $S$ by calculating the kinetic
coefficients $G_T$ and $G$.

In this paper we concentrate on the case of relatively large quantum dots,
where the effects of finite quantum level spacing $\delta$ in the dot can
be ignored.  In the limit $\delta \to 0$ the transport of electrons in
single electron transistors can be accomplished via either sequential
tunneling or inelastic cotunneling mechanisms.  The respective
contributions to the linear conductance $G$ are well known.  The
conductance as a function of the gate voltage $\phi$ shows periodic peaks
centered at points
\begin{equation}
  \label{eq:peak_centers}
  \phi_N = \frac{e}{C}\left(N + \frac{1}{2}\right),
\end{equation}
where the charging energies of the dot with $N$ and $N+1$ additional
electrons are equal.  The sequential tunneling theory
\cite{glazman:G,beenakker:G} accounts for the real events of electron
tunneling between the leads and the dot.  When the $(N+1)$-st electron
tunnels into the dot, the charging energy changes by $e(\phi_N-\phi)$.  At
low temperatures $T \ll e|\phi_N-\phi|$, the density of electrons with
energy sufficient to charge the dot is exponentially small, resulting in
the conductance peaks with exponential tails:
\begin{equation}
  \label{eq:sequential_G}
  G^{sq} = \frac{G_l G_r}{2(G_l+G_r)} 
           \frac{e(\phi-\phi_N) / T}
                {\sinh\left[e(\phi-\phi_N) / T\right]}.
\end{equation}
On the other hand, the inelastic cotunneling mechanism \cite{averin:cot}
accounts for the second-order tunneling processes when, e.g., an electron
tunnels from the left lead into the dot, and then another electron tunnels
from the dot to the right lead.  The initial and final state of such a
process have the same charge in the dot.  The Coulomb blockade only
affects the energy of the virtual state, resulting in only a power-law
suppression of the conductance at low temperatures:
\begin{equation}
  \label{eq:cotunneling_G}
  G^{co} = \frac{\pi}{3} \frac{\hbar}{e^2} G_l G_r
           \frac{T^2}{\left[ e(\phi-\phi_N) \right]^2}.
\end{equation}

The above expression formally diverges at $\phi=\phi_N$, because the
calculation in Ref.~\onlinecite{averin:cot} neglected the contribution of
the quasiparticle energies to the energy of the virtual state. To estimate
the cotunneling contribution at the center of a peak, i.e., at
$\phi=\phi_N$, one can replace the charging energy difference
$e(\phi-\phi_N)$ in the denominator of Eq.~(\ref{eq:cotunneling_G}) by the
temperature $T$.  This results in $G^{co}\sim (\hbar/e^2)G_l G_r$.  The
contribution (\ref{eq:sequential_G}) of the sequential tunneling mechanism
at the peak is $G^{sq}\sim G_l G_r/(G_l+G_r)$.  Thus, near the centers of
the peaks the conductance is dominated by sequential tunneling.

On the other hand, between the peaks the sequential tunneling contribution
(\ref{eq:sequential_G}) decays exponentially at $T\to0$, as opposed to the
relatively slow dependence $G^{co}\propto T^2$ of the cotunneling
conductance (\ref{eq:cotunneling_G}).  Therefore, at low enough
temperatures the cotunneling mechanism will dominate the conduction in the 
valleys between the peaks.  By comparing the contributions
(\ref{eq:sequential_G}) and (\ref{eq:cotunneling_G}) in the middle of a
valley, i.e., for $\phi-\phi_N=e/2C$, we find that the sequential
tunneling dominates the conductance at any gate voltage only at
temperatures $T>T_c$, where
\begin{equation}
  \label{eq:crossover_T}
  T_c\simeq\frac{E_C}{\ln[e^2/\hbar(G_l+G_r)]}.
\end{equation}
At lower temperatures, $T<T_c$, the conductance in the regions of width
\begin{equation}
  \label{eq:delta_phi}
  \Delta\phi = \frac{e}{2C}\, \frac{T}{T_c}
             \simeq\frac{T}{e}\ln\left[\frac{e^2}{\hbar(G_l+G_r)}\right]
\end{equation}
around each peak is determined by the sequential tunneling processes,
Eq.~(\ref{eq:sequential_G}), whereas outside those regions the cotunneling
processes dominate, Eq.~(\ref{eq:cotunneling_G}).  Note, that in our case
of weak tunneling, $G_l,G_r\ll e^2/\hbar$, the region (\ref{eq:delta_phi})
is wider than the thermal width of the peak $T/e$.  Thus, the cotunneling
mechanism becomes important only away from the peaks, where the
conductance is much smaller than its peak value.

The change in the transport mechanism from sequential tunneling to
cotunneling is more dramatic in the case of the thermopower.  Unlike the
conductance, the thermopower in the sequential tunneling regime does not
have the form of sharp peaks near $\phi=\phi_N$; in fact it reaches its
maximum values near the centers of the valleys between the peaks of
$G(\phi)$, Ref.~\onlinecite{beenakker:S}.  At $T<T_c$ the transport in
those regions is strongly affected by the cotunneling mechanism, which
leads to qualitative changes in both the amplitude and shape of the
Coulomb blockade oscillations of the thermopower.

\section{The thermopower in the regimes of  sequential tunneling and 
         cotunneling\label{thermopower}}

The thermopower $S$ of a system is a direct measure of the average energy
that the electrons carry during the tunneling processes. This follows from
an Onsager relation\cite{Abrikosov88} between the Peltier coefficient
\mbox{$\Pi = - \langle \xi \rangle / e$} and the thermopower:
\begin{equation}
 \label{onsager}
 S = \frac{\Pi}{T} = - \frac{\langle \xi \rangle}{e T}.
\end{equation}
The two transport mechanisms in single electron transistors,
the sequential tunneling and the cotunneling, involve
electrons with typical energies $\xi \sim E_C$ and
$\xi \sim T$, respectively. Thus the thermopower
(\ref{onsager}) is strongly affected by the crossover
between the two regimes at $T \sim T_c$.

\subsection{Sequential tunneling regime\label{sq}}

In this section we review the results for the thermopower in the
sequential tunneling regime found by Beenakker and
Staring.\cite{beenakker:S}  The sequential tunneling current is determined
by the following two elementary real transition processes: (a) an electron
tunnels between the left lead and the dot; (b) an electron tunnels between
the dot and the right lead.  Both processes are of the first order in the
perturbation parameter $G_{l,r} / (e^2/\hbar)$.  The current in the
stationary state can then be derived by means of a kinetic equation which
involves the probabilities for the system being in a certain charge state
and the tunneling rates obtained by Fermi's Golden Rule.

In the linear-response regime the current can be due to either a bias
voltage or a temperature difference between the two leads. The conductance
$G$ and the kinetic coefficient $G_T$ are found in
Refs.~\onlinecite{beenakker:G} and \onlinecite{beenakker:S}, respectively:
\begin{eqnarray}
\label{seqG}
 G^{sq} &=& 
  \frac{G_l G_r}{G_l + G_r}
  \sum_{N}^{} W_{N}^{(0)} 
  f\left(\frac{E_N-E_{N-1}}{T}\right), \\
\label{seqTI}
 G_T^{sq} &=&
  - \frac{1}{2e}\frac{G_l G_r}{G_l + G_r} 
  \sum_{N}^{} W_{N}^{(0)} 
   f\left(\frac{E_N-E_{N-1}}{T}\right)\nonumber\\
    &&\hspace{8em}\times\frac{E_N-E_{N-1}}{T}.
\end{eqnarray}
Here $E_N \equiv E_C N^2 - N e \phi$ is the electrostatic energy of the
dot containing $N$ electrons, $f(x) \equiv x / (1-e^{-x})$, and
$W_N^{(0)}$ stands for the equilibrium probability distribution of the dot
charge, $W^{(0)}_N \equiv e^{-E_N / T} / \sum_N e^{- E_N / T}$.

At low temperatures $T \ll E_C$ at most two charge states contribute
significantly to the sums in (\ref{seqG}) and (\ref{seqTI}).  One can then
neglect the exponentially small contributions of the other charge states
and obtain Eq.~(\ref{eq:sequential_G}) for the conductance $G$ as well as
an analogous expression for $G_T$,
\begin{equation}
 \label{eq:sequential_G_T}
  G^{sq}_T = \frac{G_l G_r}{4(G_l+G_r)} 
           \frac{e (\phi-\phi_N)^2 / T^2}
                {\sinh[e(\phi-\phi_N) / T]}.
\end{equation}
Using Eq.~(\ref{S}) one then obtains the
low temperature limit of the thermopower
\begin{equation}
 \label{eq:sequential_S}
 S^{sq} = \frac{\phi - \phi_N}{2 T}.
\end{equation}
Here, $\phi_N$ defines the position (\ref{eq:peak_centers})
of the conductance peak closest to $\phi$. This is the
sawtooth like behavior described in Ref.~\onlinecite{beenakker:S}.
The amplitude of the Coulomb blockade oscillations of the
thermopower is given by $S_{max}^{sq}= e/(4C T)$.

This result can be understood in terms of the average energy of tunneling
electrons. Exactly in the middle between two conductance peaks (e.g.,
$\phi = [\phi_{N-1} + \phi_{N}]/2$) the same amount $E_C$ of energy is
required to either add or remove an electron from the dot.  Therefore, the
two processes involving electrons with energies $\xi\approx E_C$ and
$\xi\approx -E_C$ contribute equally to the electronic transport, and the
average energy of tunneling electrons $\langle \xi \rangle$ vanishes.
However, if the gate potential $\phi$ is slightly changed by $\sim T/e$
towards one of the conductance peaks, one of the two charge states is
exponentially suppressed. In this case the average energy is found to be
$\langle \xi \rangle \approx \pm E_C$. This means that the thermopower
exhibits a sharp jump between two conductance peaks. If the gate voltage
is tuned further towards $\phi_N$, the average tunneling electron energy
is given by $\langle \xi \rangle \sim u_{1} \equiv E_{N+1}-E_N = e (\phi_N
- \phi)$ and thus changes linearly with $\phi$. Using this qualitative
considerations the result (\ref{eq:sequential_S}) is reproduced except for
an additional numerical factor of $1/2$. This factor $1/2$ is due to the
fact that not only electrons with energy $u_{1}$ above the Fermi level are
involved in the transport. The probability for an electron with energy
$\xi$ to tunnel into the dot is proportional to the average occupation
number $\sim e^{- \xi / T}$ in the lead and to the density of holes $\sim
e^{-(u_{1}-\xi) / T}$ in the dot.  Therefore, all electrons with energies
less than $u_{1}$ above the Fermi level have the same probability to
tunnel into the dot.  For that reason, one obtains for the average
tunneling electron energy $\langle \xi \rangle = u_{1} / 2$ which then
reproduces the result (\ref{eq:sequential_S}).

\subsection{Cotunneling regime}

As we discussed in Sec. \ref{mechanisms}, at temperatures below the
crossover temperature (\ref{eq:crossover_T}) cotunneling is the main
mechanism of transport in a single-electron transistor.  A cotunneling
process consists of two steps. First, an electron close to the Fermi level
in one lead tunnels into the dot, thus changing the energy by $\sim
u_{1}$. Since the energy is not conserved during this process, the system
is in a virtual intermediate state. A second tunneling process has to
follow which involves an electron tunneling from the dot to the other lead
and thus restoring the conservation of energy.  Since the charge of the
dot before and after a cotunneling process is the same, the electronic
transport is primarily due to electrons with energies $\xi \sim T$.
Therefore, unlike sequential tunneling, the cotunneling contribution to
the current is not exponentially suppressed at $T \ll E_C$, and it
dominates the transport in the valleys of the Coulomb blockade.  In the
following only inelastic cotunneling processes which involve two different
electrons are considered, since the elastic ones are suppressed at small
level spacing in the dot, $\delta\ll T^2/E_C$, Ref.
\onlinecite{averin:cot}.

\begin{figure}
 \resizebox{.36\textwidth}{!}{\includegraphics{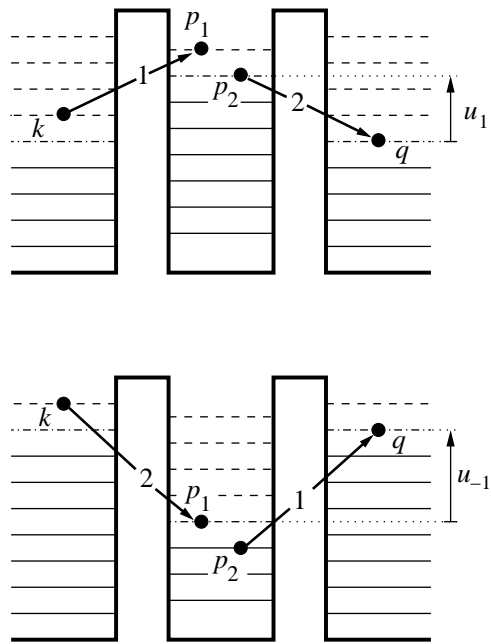}}
\caption{\label{order2} The two types of inelastic cotunneling
  processes transferring electrons from the left to the right lead.  The
  first process consists of the following two steps: an electron tunnels
  from a state $k$ in the left lead to a state $p_1$ in the dot, and then
  an electron from a state $p_2$ in the dot tunnels to state $q$ in the
  right lead.  The second process transfers electrons in the opposite
  order: it starts with an electron tunneling from the dot to the right
  lead, and finishes with another electron tunneling from the left lead
  into the dot.  The energies of the virtual states in the two types of
  processes are affected by the electrostatic energies $u_1$ and $u_{-1}$
  required to either increase or decrease the charge of the dot by that of
  one electron.  The dash-dotted lines show the positions of the Fermi
  level in the leads and the dot.}
\end{figure}

The cotunneling contribution to the current is of the second order in the
small parameter $G_{l,r}/(e^2/\hbar)$ because two coherent tunneling
processes have to take place.  For a given number $N$ of electrons in the
dot there are two types of second order processes which can transfer
electron from one lead to the other, see Fig. \ref{order2}. The current
then reads
\begin{eqnarray}
\label{cotcurrent}
 I &=& - \frac{2 \pi}{\hbar} e \sum_{k,p_1,p_2,q}
          [n_k (1-n_{p_1}) n_{p_2} (1-n_q) \nonumber\\
    && - n_q (1-n_{p_2}) n_{p_1} (1-n_k)]
    \delta (\xi_k-\xi_{p_1}+\xi_{p_2}-\xi_q)\nonumber\\
    &&\times\left| \frac{t_{k,p_1} t_{p_2,q}}{\xi_k-\xi_{p_1} - u_1}
  - \frac{t_{k,p_1} t_{p_2,q}}{\xi_k-\xi_{p_1} + u_{-1}}\right|^2, 
\end{eqnarray}
where $n_k$, $n_{p_1}$, $n_{p_2}$ and $n_q$ are the occupation numbers of
the respective states; $u_{\pm1} \equiv E_{N\pm1} - E_N$, and $N$ is
determined by minimizing the electrostatic energy of the dot for a given
potential $\phi$.  Note that the first product of the occupation numbers
corresponds to the current from the left to the right lead, whereas the
second one accounts for the current in the opposite direction.

If the bias voltage $V$ is zero, and all three electrodes are
at the same temperature $T$, the current
from the left lead to the right lead cancels with 
the one flowing into the opposite direction.
This can most easily be seen by exchanging 
$k\leftrightarrow q$ and $p_1 \leftrightarrow p_2$ in one 
of the products of occupation numbers which is possible 
because of the apparent left-right symmetry of the system.

The cotunneling contribution $G^{co}$ to the conductance 
can be derived by linearizing Eq.~(\ref{cotcurrent}) with respect
to the bias voltage $V$. After replacing the sums by integrals
and the occupation numbers by Fermi functions, the conductance
can be presented as a single integral,
\begin{equation}
\label{eq:G_co_integral}
 G^{co} = \frac{\hbar G_l G_r}{8 \pi e^2 T} \int
          \frac{\xi^2}{\sinh^2 \frac{\xi}{2T}}
          \left| \frac{1}{\xi - u_1} - 
                 \frac{1}{\xi + u_{-1}} \right|^2 d\xi.
\end{equation}
This expression formally diverges at $\xi = \pm u_{\pm 1}$; we will
discuss the proper regularization procedure in
Sec.~\ref{subsec:low_T_thermopower}. At low temperatures $T \ll u_1 ,
u_{-1}$ the contributions of the regions near the singularities are
exponentially small, and one can evaluate $G^{co}$ by neglecting the
quasiparticle energies $\xi \sim T$ in the denominators.  Then the
conductance $G^{co}$ is found \cite{averin:cot} to be
\begin{equation}
\label{cotG}
 G^{co} = \frac{\pi \hbar}{3 e^2}\, G_l G_r T^2 \left(
     \frac{1}{u_1} + \frac{1}{u_{-1}} \right)^2.
\end{equation} 
If the potential $\phi$ is significantly closer to the peak at $\phi_N$
than to $\phi_{N-1}$ the term $1/u_{-1}$ can be neglected compared to
$1/u_{1}$, and Eq.~(\ref{eq:cotunneling_G}) is reproduced.

In order to calculate the kinetic coefficient $G_T$ we assume a slightly
higher temperature $T + \Delta T$ in the left lead compared to the dot and
the right lead.  Then the current (\ref{cotcurrent}) is linearized with
respect to the small temperature difference $\Delta T \ll T$.  After
replacing the sums by integrals and occupation numbers by Fermi functions,
the following expression for $G_T$ is obtained:
\begin{widetext}
\begin{eqnarray}
\label{eq:GcoT}
 G^{co}_T &=& \frac{\hbar G_l G_r}{2 \pi e^3}
  \int d\xi_k \, d\xi_{p_1} \, d\xi_{p_2} \, d\xi_{q} \; \delta 
  (\xi_k - \xi_{p_1}+\xi_{p_2}-\xi_q)
  \left| \frac{1}{\xi_k - \xi_{p_1} - u_1} - 
  \frac{1}{\xi_k - \xi_{p_1} + u_{-1}} \right|^2 \nonumber \\
 &&\times \, \frac{\xi_k}{T} \; \frac{d n(\xi_k)}{d \xi_k}
   \left\{ [1-n(\xi_{p_1})] n(\xi_{p_2}) [1-n(\xi_q)]
          + n(\xi_{p_1}) [1-n(\xi_{p_2})] n(\xi_q) \right\}.
\end{eqnarray}
\end{widetext}
The same expression is obtained in a more careful treatment where the
temperature of the dot is not necessarily equal to the temperature of the
right lead.  Three of the four integrals can be calculated exactly,
resulting in the following expression for $G_T$
\begin{equation}
\label{eq:G_co_T_integral}
 G^{co}_T = - \frac{\hbar G_l G_r}{16 \pi e^3} \frac{1}{T^2}
  \int \frac{\xi^3}{\sinh^2 \frac{\xi}{2T}}
  \left| \frac{1}{\xi - u_1} - 
  \frac{1}{\xi + u_{-1}} \right|^2 d\xi.
\end{equation}
As it was the case for the conductance $G$, the main
contribution to $G_T$ comes from energies $\xi$ of the order
of $T \ll u_1 , u_{-1}$.
However, setting those terms in the denominators to
zero yields a vanishing
$G_T$ as the integrand in (\ref{eq:G_co_T_integral}) becomes an
odd function in $\xi$.
The first non-vanishing contribution to $G_T$ is obtained
by expanding the fractions in the integral up to first order
in $\xi / u_1$.
The final result for $G_T^{co}$ can be written as
\begin{equation}
\label{cotTI}
 G_T^{co} = - \frac{4 \pi^3}{15} \frac{\hbar}{e^3} 
  \, G_l G_r T^3
 \left( \frac{1}{u_1} + \frac{1}{u_{-1}} \right)
 \left( \frac{1}{u_1^2} - \frac{1}{u_{-1}^2} \right).
\end{equation}

Using the expression (\ref{S}) for the thermopower
we find from (\ref{cotG}) and (\ref{cotTI}) 
\begin{equation}
\label{cotS}
 S^{co} = \frac{4 \pi^2}{5} \frac{T}{e^2} \left( \frac{1}{\phi - \phi_{N}} 
     + \frac{1}{\phi - \phi_{N-1}} \right).
\end{equation}
This expression corresponds to the potential range between two conductance
peaks $\phi_{N-1} < \phi < \phi_{N}$ with the exception of the values of
$\phi$ very close to the peaks, namely $(\phi_N - \phi), (\phi -
\phi_{N-1}) \gg T/e$.

The result (\ref{cotS}) can be understood in terms of the average energy
of tunneling electrons, Eq.~(\ref{onsager}). The transport in the
cotunneling regime is mainly due to the electrons with energies $\xi \sim
\pm T$.  In the low temperature limit $T \ll u_{1}$ the tunneling
probability from a state with energy $\xi$ into an intermediate state
with energy $\xi_p$ is proportional to
\begin{equation}
\label{cotprobability}
 w(\xi) \propto \frac{1}{(\xi_p - \xi + u_{1})^2}
        \simeq \frac{1}{u_{1}^2} 
             \left( 1 + 2 \frac{\xi - \xi_p}{u_{1}} \right).
\end{equation}
This expression clearly shows that the tunneling probability $w(\xi)$ is
enhanced for electrons above the Fermi level $\xi > 0$, and therefore the
average energy $\langle \xi \rangle$ will not vanish. In fact, with a
typical energy $\xi \sim T$ for electrons involved in the tunneling and
the probability (\ref{cotprobability}) we find for the average energy
$\langle \xi \rangle \sim T^2/u_{1}$.  Using the relation (\ref{onsager}),
the cotunneling contribution to the thermopower (\ref{cotS}) is reproduced
in the correct order of magnitude.

\section{The thermopower at arbitrary gate voltage\label{general}}

In the previous section expressions for the thermopower in both the
sequential tunneling and cotunneling regime were presented. The sequential
tunneling result is given by Eq. (\ref{eq:sequential_S}). It dominates the
thermopower for gate potentials $\phi$ close to the positions of the
conductance peaks $\phi_N$.  On the other hand, the cotunneling result
(\ref{cotS}) gives the correct description of the thermopower between the
conductance peaks $\phi_{N-1} < \phi < \phi_N$ if the temperature $T$ is
below the crossover temperature $T_c$.  In this section we find the
combined contributions of both transport mechanisms and obtain an
expression for the thermopower valid at any gate voltage.

The current through the quantum dot is the sum of the two contributions,
sequential tunneling and cotunneling.  Therefore using
Eq.~(\ref{eq:linear_response}) and (\ref{S}) we find for the thermopower
\begin{equation}
\label{eq:general_S}
S = \frac{G_T^{sq} + G_T^{co}}{G^{sq} + G^{co}}.
\end{equation}
The well known\cite{beenakker:G,beenakker:S} sequential tunneling
contributions $G^{sq}$ and $G_T^{sq}$ are valid for the entire potential
range. However, the cotunneling results (\ref{cotG}) and (\ref{cotTI}) are
only valid away from the centers of the conductance peaks, i.e. $(\phi_N -
\phi), (\phi - \phi_{N-1}) \gg T/e$, and diverge close to the peaks.  In
the following we will show how to correctly regularize these divergences.
First we discuss the thermopower (\ref{eq:general_S}) in the limit of very
low temperatures $T \ll T_c$.  Then we present the more general result
valid for higher temperatures.

\subsection{The low temperature thermopower in the vicinity
 of a conductance peak\label{subsec:low_T_thermopower}}

In the regime of very low temperatures $T \ll T_c$ there are at most two
charge states, e.g., $N$ and $N+1$ electrons in the dot, that contribute
significantly to the total current while the contributions of all other
states are exponentially suppressed. The crossover between the sequential
tunneling and the cotunneling regimes takes place at a gate potential
close to the conductance peak $|\phi - \phi_N| \sim \Delta \phi$, where
$\Delta \phi$ is given by Eq.~(\ref{eq:delta_phi}). In order to find the
behavior of the thermopower (\ref{eq:general_S}) in this potential range
we can neglect terms $\sim 1/u_{-1}$ occurring in the current
(\ref{cotcurrent}).

To regularize the singularities at $\xi_k - \xi_{p_1} = u_1$ in the
expression (\ref{cotcurrent}) for the cotunneling current we follow
the approach presented in Ref.~\onlinecite{averin:G} and add a small
imaginary part $\Gamma$ to the energy of the intermediate virtual
states in Eq.~(\ref{cotcurrent}),
\begin{eqnarray}
\label{gencurrent}
 I &=& - \frac{2 \pi}{\hbar} e \sum_{k,p_1,p_2,q}[ n_k
 (1-n_{p_1}) n_{p_2} (1-n_q) \nonumber\\
    && - n_q (1-n_{p_2}) n_{p_1} (1-n_k)]\,
       \delta(\xi_k-\xi_{p_1}+\xi_{p_2}-\xi_q)
\nonumber \\ & &  \times 
\left| \frac{t_{k,p_1}
 t_{p_2,q}}{\xi_k-\xi_{p_1} - u_1 - i\Gamma} \right|^2.
\end{eqnarray}

\begin{figure}
 \resizebox{.38\textwidth}{!}{\includegraphics{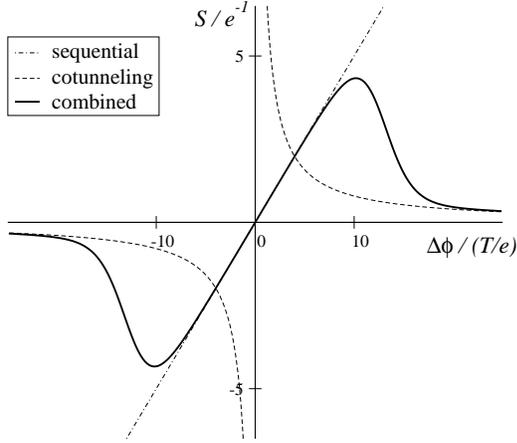}}
\caption{\label{s_lowT}The thermopower $S$ of a single electron transistor
  in the vicinity of a conductance peak.  The distance from the peak
  $\Delta\phi$ is assumed to be small compared to the distance to the next
  peak, $\Delta \phi\ll e/C$.  The solid curve is calculated for
  $G_l+G_r=10^{-3}\times\frac{2\pi e^2}{\hbar}$ by substituting the
  expressions (\ref{eq:sequential_G}), (\ref{eq:sequential_G_T}),
  (\ref{eq:result_G_co}), and (\ref{eq:result_G_co_T}) into
  Eq.~(\ref{eq:general_S}).  The asymptotics of cotunneling and sequential
  tunneling are shown by dashed and dash-dotted lines, respectively.}
\end{figure}

In the limit $\Gamma \to 0$ this expression can be divided into a
large part $\sim 1/\Gamma$ and a smaller one that is independent of
$\Gamma$ using the following procedure:
\begin{eqnarray}
\label{recipe}
 \int dE \frac{f(E)}{E^2 + \Gamma^2} &=& \int dE \frac{f(0)}{E^2 +
 \Gamma^2} + \int dE \frac{f(E)-f(0)}{E^2 + \Gamma^2} \nonumber \\
 &\to &  \frac{\pi}{\Gamma} \, f(0) +  \int dE
 \frac{f(E)-f(0)}{E^2},
\end{eqnarray}
where the last integral is to be understood as the principal value.  As we
show in the Appendix~\ref{app:seq_limit}, the first term corresponds to
the sequential tunneling.  In particular, the results
(\ref{eq:sequential_G}) and (\ref{eq:sequential_G_T}) are restored in the
linear response regime using the scheme (\ref{recipe}) and
Eq.~(\ref{eq:Gamma_N_equilibrium}) for $\Gamma$. The second term
represents the correct non-divergent expression for the cotunneling.  We
apply the scheme (\ref{recipe}) by rewriting the cotunneling conductance
(\ref{eq:G_co_integral}) and the kinetic coefficient
(\ref{eq:G_co_T_integral}) according to the following rule:
\begin{eqnarray}
\label{rule}
\int dE \frac{f(E)}{E^2} &\to& \int dE \frac{f(E) - f(0)}{E^2}
  \nonumber\\
 &=&  \int dE \frac{f(E) + f(-E) - 2 f(0)}{2E^2}.
\end{eqnarray}
This regularization procedure coincides with the one discussed in
Ref.~\onlinecite{schoeller} for the case of multichannel tunneling
junctions.  The result for $G^{co}$ and $G_T^{co}$ then reads
\begin{eqnarray}
\label{eq:result_G_co}
 G^{co} &=& \frac{\hbar}{4\pi e^2} \; G_l G_r \; {\cal
 F}\left[ \frac{e (\phi - \phi_N)}{2T} \right] ,\\
\label{eq:result_G_co_T}
 G_T^{co} &=& \frac{\hbar}{4\pi e^3} \; G_l G_r \; {\cal
 F}_T \left[ \frac{e (\phi - \phi_N)}{2T} \right],
\end{eqnarray}
where the functions ${\cal F}$ and ${\cal F}_T$ are defined as
\begin{eqnarray}
\label{eq:F}
 {\cal F}(x) &\equiv& |x| \int\limits_0^\infty \frac{dz}{z^2} \left(
 \frac{(1+z)^2}{\sinh^2[x(1+z)]} \right.\nonumber\\ & &\left. \hspace{3em}
   +\frac{(1-z)^2}{\sinh^2[x(1-z)]}-\frac{2}{\sinh^2[x]} \right),\\
\label{eq:F_T}
 {\cal F}_T(x) &\equiv& x |x| \int\limits_0^{\infty} \frac{dz}{z^2}
 \left( \frac{(1+z)^3}{\sinh^2[x(1+z)]} 
   \right.\nonumber\\ & &\left. \hspace{3em}
   +\frac{(1-z)^3}{\sinh^2[x(1-z)]} - \frac{2}{\sinh^2[x]} \right).
\end{eqnarray}
The final result for the thermopower is then obtained by substituting the
sequential tunneling results (\ref{eq:sequential_G}) and
(\ref{eq:sequential_G_T}) and the regularized cotunneling results
(\ref{eq:result_G_co}) and (\ref{eq:result_G_co_T}) into
Eq.~(\ref{eq:general_S}).  The dependence of the thermopower on the gate
voltage obtained from those expressions is illustrated in
Fig.~\ref{s_lowT}.

As can be seen in Fig.~\ref{s_lowT}, the thermopower is dominated by
the sequential tunneling contribution for small $\Delta \phi = \phi
-\phi_N$.  After it reaches its maximum value $S_{max}$, the
thermopower falls off sharply to merge with the cotunneling result
(\ref{cotS}) for large $\Delta \phi$.  To understand this behavior one
should notice that the crossover between the sequential tunneling and
the cotunneling occurs at different values $\Delta \phi_1$ and
$\Delta\phi_2$ for the conductance $G$ and for $G_T$.  The two
crossover values of the gate voltage can be estimated by setting
$G^{sq} = G^{co}$ and $G^{sq}_T = G^{co}_T$.  At small conductance $g
\equiv \hbar(G_l+G_r) / (2 \pi e^2)\ll 1$ the Eqs.~(\ref{eq:sequential_G}),
(\ref{cotG}), (\ref{eq:sequential_G_T}) and (\ref{cotTI}) result in
\begin{eqnarray}
\label{eq:phi_1}
 \frac{e \Delta \phi_1}{T} \simeq \ln \left[ \frac{1}{g} \left( \ln
 \frac{1}{g} \right)^3 \right], \\
\label{eq:phi_2}
 \frac{e \Delta \phi_2}{T} \simeq \ln \left[ \frac{1}{g} \left( \ln
 \frac{1}{g} \right)^5 \right].
\end{eqnarray}
The thermopower at $\Delta \phi < \Delta \phi_1$ is given by the
sequential tunneling result $S \simeq \Delta \phi /2T$.  In the narrow
range $\Delta \phi_1 < \Delta \phi < \Delta \phi_2$ the conductance is
already dominated by the cotunneling, while the main contribution to $G_T$
in the numerator of (\ref{eq:general_S}) is still due to the sequential
tunneling.  Therefore we find a steeply descending thermopower $S\propto
\exp[-e \Delta \phi / T]$ in that region.  Eventually, at $\phi > \phi_2$
both the numerator and the denominator in Eq.~(\ref{eq:general_S}) are
governed by the cotunneling contribution, and $S \sim T / (e \Delta
\phi)$.  Therefore the maximum of the thermopower is reached at $\Delta
\phi\simeq \Delta \phi_1$ and can be obtained using the sequential
tunneling expression (\ref{eq:sequential_S}):
\begin{equation}
\label{eq:S_max}
S_{max} \simeq \frac{1}{2e} \ln \frac{1}{g}.
\end{equation}
To find the value of the thermopower after the exponential
fall-off, one can substitute $\Delta \phi_2$ into the cotunneling
expression (\ref{cotS}), which gives $S\simeq(4\pi^2/5)e^{-1}
\ln^{-1}\frac{1}{g} \ll S_{max}$.

\subsection{The thermopower at arbitrary temperatures}

In Sec.~\ref{subsec:low_T_thermopower} we studied the thermopower
at very low temperatures $T \ll T_c$. In this regime the
thermopower in the valleys between the conductance peaks
centered at $\phi = \phi_N$ is described by the simple
cotunneling result (\ref{cotS}); the sequential tunneling
contributes only in narrow regions around $\phi = \phi_N$.
Another interesting regime is that of the temperatures
of the order of $T_c$, where one can explore the crossover
from the sawtooth behavior (\ref{eq:sequential_S}) to
the correct low-temperature limit.

\begin{figure}[b]
\vspace{\baselineskip}
 \resizebox{.38\textwidth}{!}{\includegraphics{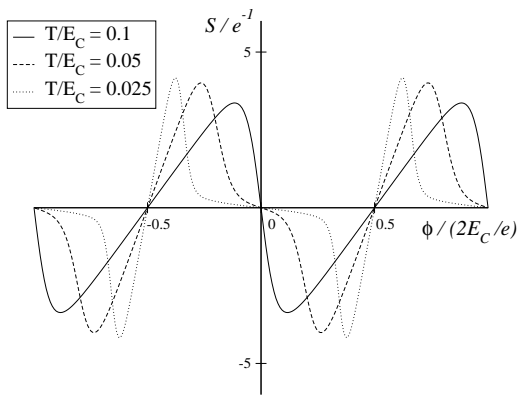}}
\caption{\label{thermo} The thermopower of a single electron transistor at
  different temperatures.  Below the crossover temperature $T_c$ the
  thermopower in the valleys of Coulomb blockade is suppressed, and the
  shape of the oscillations changes from the sawtooth (solid line) typical
  of sequential tunneling to the low-temperature behavior shown by dotted
  line.  The curves were calculated for $G_l+G_r=10^{-3}\times\frac{2\pi
    e^2}{\hbar}$ by substituting Eqs.~(\ref{seqG}), (\ref{seqTI}),
  (\ref{final_G_co}), and (\ref{final_G_co_T}) into
  Eq.~(\ref{eq:general_S}).}
\end{figure}

The crossover takes places in the valleys between the peaks,
where more than two charge states give comparable contributions
to the current through the device. Thus we present the
total cotunneling current as a weighted sum of the
contributions of states with a different number $N$ of
electrons in the dot, $I=\sum W_N^{(0)} I^{co}_N$, where
$I_N^{co}$ is given by
\begin{widetext}
\begin{eqnarray}
\label{gen_cot_current}
 I^{co}_N &=& - \frac{2 \pi}{\hbar} e \sum_{k,p_1,p_2,q}
            \left[ n_k (1-n_{p_1}) n_{p_2} (1-n_q) 
            - n_q (1-n_{p_2}) n_{p_1} (1-n_k) \right]
         \delta (\xi_k-\xi_{p_1}+\xi_{p_2}-\xi_q)
   \nonumber \\
 & &\hspace{6em} \times 
  \left| \frac{t_{k,p_1} t_{p_2,q}}{\xi_k-\xi_{p_1} - E_{N+1} + E_N}
  - \frac{t_{k,p_1} t_{p_2,q}}{\xi_k-\xi_{p_1} + E_{N-1} - E_N}\right|^2.
\end{eqnarray}
Note that one has to keep terms $\sim 1/u_{-1} = 1/(E_{N-1}-E_N)$
if the expression is to be valid not only close to the peaks
at $\phi_N$ but also in the valleys between the peaks.
Similarly to the discussion in the previous section,
we linearize the current with
respect to a small bias voltage $V$ and to a small temperature
difference $\Delta T$, yielding $G$ and $G_T$. The cotunneling
contributions have then to be regularized applying the
rule (\ref{rule}) to each of the divergent terms.
The result is
\begin{eqnarray}
\label{final_G_co}
 G^{co} &=& \frac{\hbar}{e^2} \frac{G_l G_r}{4 \pi} 
  \sum_{N=-\infty}^{\infty} \left[ ( W_{N-1}^{(0)} + W_N^{(0)} ) 
  {\cal F} \left( \frac{E_N-E_{N-1}}{2T} \right)
  - \frac{4TC}{e^2} 
  ( W_{N-1}^{(0)} - W_N^{(0)} ) {\cal F}^*
  \left( \frac{E_N-E_{N-1}}{2T} \right) \right], \\
\label{final_G_co_T}
 G^{co}_T &=& - \frac{\hbar}{e^3} \frac{G_l G_r}
  {4 \pi} \sum_{N=-\infty}^{\infty} 
  \left[ ( W_{N-1}^{(0)} + W_N^{(0)} ) {\cal F}_T
  \left( \frac{E_N - E_{N-1}}{2T} \right)
  - \frac{4TC}{e^2} 
  (W_{N-1}^{(0)} - W_N^{(0)} ){\cal F}_T^*
  \left( \frac{E_N - E_{N-1}}{2T} \right) \right],
\end{eqnarray}
where the functions ${\cal F}^*$ and ${\cal F}_T^*$
are defined by
\begin{eqnarray}
\label{eq:F^*}
 {\cal F}^*(x) &\equiv& x |x| \int\limits_0^\infty \frac{dz}{z}
  \left( \frac{(1+z)^2}{\sinh^2[x(1+z)]} - 
  \frac{(1-z)^2}{\sinh^2[x(1-z)]}\right), \\
\label{eq:F_T^*}
 {\cal F}_T^*(x) &\equiv& x^2 |x| \int\limits_0^{\infty} 
 \frac{dz}{z} \left( \frac{(1+z)^3}{\sinh^2[x(1+z)]} - 
 \frac{(1-z)^3}{\sinh^2[x(1-z)]} \right).
\end{eqnarray}
\end{widetext}
The nondivergent cotunneling contributions $G^{co}$ and $G_T^{co}$
together with the well known sequential tunneling results (\ref{seqG}) and
(\ref{seqTI}) for $G^{sq}$ and $G_T^{sq}$ then give the thermopower
according to Eq.~(\ref{eq:general_S}).  This expression for the
thermopower $S$ is the final result of the calculation.  It is valid at
any gate voltage $\phi$.  As the temperature changes, this result shows
the crossover from the sawtooth shape of the Coulomb blockade oscillations
of the thermopower at $T\gg T_c$ to the low-temperature behavior
discussed in Sec.~\ref{subsec:low_T_thermopower} at $T\ll T_c$, see
Fig.~\ref{thermo}.  The low-temperature curve of $S(\phi)$ shown by dotted
line in Fig.~\ref{thermo} is qualitatively similar to the experimentally
measured thermopower of Ref.~\onlinecite{dzurak:exp}.

\section{Summary}

We have studied the thermopower of a single electron transistor based on a
quantum dot weakly coupled to two leads, Fig.~\ref{system}.  The transport
through the dot is governed by two different electronic transport
mechanisms: the sequential tunneling and the cotunneling.  At temperatures
above the crossover temperature $T_c$ given by Eq.~(\ref{eq:crossover_T})
the sequential tunneling dominates.  The thermopower in this regime was
studied in Ref.~\onlinecite{beenakker:S}.  At temperatures below $T_c$ the
cotunneling mechanism gives the main contribution to the transport in the
valleys between the Coulomb blockade peaks of conductance.  This changes
the shape of the Coulomb blockade oscillations of the conductance
dramatically, Fig.~\ref{thermo}.  We derived a single expression for the
thermopower at arbitrary temperature including both contributions.  It is
given by Eq.~(\ref{eq:general_S}) in combination with Eqs.~(\ref{seqG}),
(\ref{seqTI}), (\ref{final_G_co}), and (\ref{final_G_co_T}).  As can be
seen in Fig.~\ref{thermo}, the resulting thermopower significantly changes
its shape from the sawtooth behavior for temperatures above $T_c$ to the
low-temperature behavior at $T\ll T_c$.  In the regime of low temperatures
the amplitude of the thermopower becomes temperature independent,
Eq.~(\ref{eq:S_max}).  Near the maximum of the thermopower the transport
is determined by the balance of the two mechanisms.  The shape of the
thermopower peaks is studied in detail in
Sec.~\ref{subsec:low_T_thermopower} and illustrated in Fig.~\ref{s_lowT}.

\begin{acknowledgments}
  This research was supported by the NSF Grant No.~DMR-9974435 and by the
  Sloan foundation.  We are also grateful to B.L. Altshuler and A.V.
  Andreev for valuable discussions.
\end{acknowledgments}

\appendix*
\section{Sequential tunneling limit\label{app:seq_limit}}

In this appendix we show how the sequential tunneling
contribution $I^{sq}$ can be extracted from the current
(\ref{gencurrent}) by taking the $\Gamma \to 0$ limit
and keeping only large terms $\sim 1/\Gamma$. Using
the scheme (\ref{recipe}) the fraction
$1/[(\xi_k - \xi_{p_1} - u_1)^2 + \Gamma^2]$ is
replaced by $\pi \delta (\xi_k - \xi_{p_1} - u_1) / \Gamma$.
The result can be presented as
\begin{equation}
\label{eq:current_Gamma_to_0}
 I^{sq} = - e \, \frac{R^l_{N \to N+1} \; R^r_{N+1 \to N} -
                  R^l_{N+1 \to N} \; R^r_{N \to N+1}}
                 {2 \, \Gamma / \hbar},
\end{equation}
where the tunneling rates $R$ are given by Fermi's Golden
Rule. For the tunneling between the left lead and the
dot they are
\begin{eqnarray*}
 R^l_{N \to N+1} &=& \frac{2 \pi}{\hbar} \sum_{k,p}
  |t_{k,p}|^2 \, n_k \left(1-n_p\right) \, 
   \delta (\xi_k - \xi_p - u_1),\\
 R^l_{N+1 \to N} &=& \frac{2 \pi}{\hbar} \sum_{k,p}
  |t_{k,p}|^2 \, n_p \left(1-n_k\right) \, 
   \delta (\xi_k - \xi_p - u_1),
\end{eqnarray*}
and $R^r$ can be obtained from the expressions for $R^l$ by replacing the
indices $k \to q$.

On the other hand, in the sequential tunneling theory
\cite{beenakker:G} the current is presented as
\begin{eqnarray}
\label{eq:current_sequential}
I^{sq} &=& - e \left(W_N \, R^l_{N \to N+1} - 
                   W_{N+1} \, R^l_{N+1 \to N}\right)
\nonumber\\
       &=& - e \left(W_{N+1} \, R^r_{N+1 \to N} - 
                   W_N \, R^r_{N \to N+1}\right),
\end{eqnarray}
where $W_N$ represents the probability of the system
being in a state with $N$ electrons in the dot.
These two expressions for the current $I^{sq}$ arise from
the condition that the current through the left barrier
equals the current through the right barrier.
Employing this equality the current (\ref{eq:current_sequential})
can be rewritten as
\begin{equation}
 I^{sq} = - e W_N \frac{R^l_{N \to N+1} R^r_{N+1 \to N} -
                  R^l_{N+1 \to N} R^r_{N \to N+1}}
                 {R^l_{N+1 \to N} + R^r_{N+1 \to N}}.
\end{equation}
This is the same relation as Eq.~(\ref{eq:current_Gamma_to_0})
if we make the identification
\begin{equation}
\label{eq:Gamma_N}
 \Gamma = \frac{\hbar}{2} 
  \frac{R^l_{N+1 \to N} + R^r_{N+1 \to N}}{W_N}.
\end{equation}
This expression coincides with the one proposed in
Ref.~\onlinecite{averin:G},
\begin{equation}
 \Gamma = \frac{\hbar}{2} 
      \left( R^l_{N+1 \to N} + R^r_{N+1 \to N} +
             R^l_{N \to N+1} + R^r_{N \to N+1} \right),
\end{equation}
if one takes into account the balance equation
(\ref{eq:current_sequential}) and the condition $W_N + W_{N+1} = 1$. Our
derivation shows that replacing the denominator in (\ref{gencurrent}) with
a $\delta$-function reproduces the result of the sequential tunneling even
if $eV$ and/or $\Delta T$ are not small compared to the temperature $T$.
In Sec.  \ref{subsec:low_T_thermopower} we use this result in the linear
response regime, $eV, \Delta T \ll T$, to conclude that this approximation
reproduces the results (\ref{eq:sequential_G}) for $G^{sq}$ and
(\ref{eq:sequential_G_T}) for $G_T^{sq}$. In this case the $\Gamma$ is
determined by the equilibrium rates and can be calculated by replacing the
occupation numbers with Fermi functions,
\begin{equation}
\label{eq:Gamma_N_equilibrium}
 \Gamma  = \frac{\hbar}{2e^2} (G_l+G_r) \, u_1 \, 
             \coth \left[ \frac{ u_1}{2T} \right],
\end{equation}
in agreement with Ref.~\onlinecite{averin:G}.

\end{document}